# Negative Thermal expansion of pure and doped Graphene

Sarita Mann, Ranjan Kumar and V.K. Jindal[1]

Department of Physics, Panjab University Chandigarh-160014.

Graphene and its derivatives distinguish themselves for their large negative thermal expansion even at temperatures as high as 1000K. The linear thermal expansion coefficients (LTEC) of two-dimensional honeycomb structured pure graphene and B/N doped graphene are analyzed using ab initio density functional perturbation theory (DFPT) employed in VASP software under quasiharmonic approximation. One of the essential ingredients required is the phonon frequencies for a set of points in the Brillouin zone and their volume dependence. These were obtained from the dynamical matrix which was calculated using VASP code in interface with phonopy code. In particular, the transverse acoustic modes (ZA) behave drastically differently as compared to planer modes and so also their volume dependence. Using this approach firstly thermal expansion for pure graphene is calculated. The results agree with earlier calculations using similar approach. Thereafter we have studied the effect of boron and nitrogen doping on LTEC. The LTEC of graphene is found to be negative in the entire range of temperature under study (0-1000K) and its value at room temperature (RT) is around $-3.26 \times 10^{-6} K^{-1}$. The value of LTEC at RT becomes more negative with B/N doping in graphene. In order to get an insight into the cause of negative thermal expansion, we have computed the contribution of individual phonon modes of vibration. We notice that it is principally the ZA modes which are responsible for negative thermal expansion. It has been concluded that transverse mode in 2D hexagonal lattices have an important role to play in many of the thermodynamical properties of 2D structures. We have extended the study to calculate the LTEC of h-BN sheet also.

**Keywords**- Graphene, h-BN, density functional perturbation theory, phonons, doping, thermal expansion, Grüneisen parameter

## 1. INTRODUCTION

The first successful experimental synthesis of graphene [1], a 2D thermodynamically stable structure of carbon, created a great deal of research interest in it. In graphene $sp^2$ hybridized carbon atoms are arranged in honeycomb lattice structure. It has attracted great scientific interest mainly because of its unusual electronic properties In fact it has been identified as a very promising candidate as future application material based on its chemical and mechanical stability along with ballistic transport at room temperature (RT) [2]. Its charge carriers exhibit very high intrinsic mobility. Its special electronic transport which is governed by a Dirac-like-equation makes it act as a bridge between the condensed matter physics and electrodynamics [3-5].There has been a lot of research ongoing to study and modify electronic and optical properties of graphene by doping with a variety of atoms. We in our previous research [6-8] have studied the band gap modification and optical behavior by optical behavior by replacing some percentage of the carbon atoms with nitrogen/boron atom dopants.

---

[1]Email for correspondence: Jindal@pu.ac.in

Equally important aspect of graphene is its unique thermal properties. Its thermal conductivity is unusually high [9] which is a subject of intensive research. Heat flow hindrance in miniature components is a matter of concern in the micro and nano-electronics. There is a lot of research ongoing in this area. In addition to this, graphene also has unusual negative linear thermal expansion coefficient (LTEC) as reported by several authors [10-15]. Mounet et al. [10] have made a detailed study of LTEC of carbon materials including graphene using first-principles based quantum espresso code under quasi harmonic approximation (QHA). They report the LTEC for graphene upto temperatures as high as 2500K with a RT value around $-3.6 \times 10^{-6}$ $K^{-1}$. Calculations made by Monte Carlo simulations [11] estimated the average value of LTEC of graphene to be $-4.8 \times 10^{-6}$ $K^{-1}$ between the temperatures ranges 0–300 K and estimated negative behavior only up to 900 K. We also notice a calculation [12] determining the LTEC of graphene using non-equilibrium green's function with and without substrate. They report large negative values of LTEC at low temperature region and small positive values at high-temperature region, with a RT value of $-6 \times 10^{-6}$ $K^{-1}$. Thereafter Sevik [13] has reported similar results as [10] for graphene using DFPT implemented in VASP. However none of the above theoretical calculations bring out clearly the origin of the negative thermal expansion though it remains a feature of 2D systems which is also found to be reported by Kim et al [14] who observe it for graphyne too. Recently, Jiang *et al.* [15] have reviewed the significance of flexural mode of graphene. They report that flexural mode is responsible for high negative contribution to thermal expansion of graphene by studying the contribution of individual phonon modes to thermal expansion. Michel et al [16] have made a detailed analysis on the role of anharmonic phonons on 2D crystals reporting negative thermal expansion whenever flexural modes are involved. They [17] also report selective contribution from various acoustic modes to thermal expansion in graphene and introduce finite size to check the divergent Grüneisen parameter from flexural modes.

Experimental studies of thermal expansion of graphene have also been reported [18-20]. Bao *et al.* [18] have measured the change in length of suspended graphene sample and estimated the LTEC of graphene in a narrow range of temperature (300-400K), which remains negative only upto 350K with a RT value of -$7 \times 10^{-6}$ $K^{-1}$. Yoon *et al.* [19] have measured temperature dependent Raman shift in G band of graphene and analytically estimated the LTEC in temperature range of (200-400) K, with RT value of $(-8 \pm 0.7) \times 10^{-6}$ $K^{-1}$. Similarly, Pan W. et al. [20] also report useful data for thermal expansion in the high temperature range from RT to above 1000 K from the measured Raman shifts of G and 2D peaks in graphene. Most of the theoretical estimates [10-17] underestimate the experimental observations.

The h-BN (hexagonal-boron nitride) sheet is another two-dimensional (2D) material which has attracted great attention in last few years due to its interesting application related properties [21, 22]. The h-BN crystal structure is similar to graphene as boron and nitrogen atoms are arranged in $sp^2$ bonded 2D hexagonal structure. In the h-BN sheet, nitrogen and boron atoms are attached by strong covalent bonds. Despite the polar nature of the BN bond in contrast to graphene which lead to significant electronic structural changes, its elastic properties remain similar. h-BN distinguishes itself also because of a large band gap of 5-6eV as compared to zero band gap in graphene[23]. It also shows high thermal stability, large dielectric breakdown and enhanced chemical inertness compared to graphene. On account of these properties, it becomes a useful material for applications in industries. Therefore, it follows naturally that the unique electronic properties are supplemented by a study of their thermodynamic and phonon related properties as well, only then the full potential in the applications of graphene or doped graphene or BN sheet is exploited. Sevik [13] has also studied LTEC of h-BN sheet using VASP software and reported negative thermal expansion with almost double the RT value as compared to graphene. Thomas *et al.* [24] have studied the LTEC of h-BN sheet using molecular dynamics and shows negative LTEC at low temperature.

Negative LTEC is crucial, as it is reported in almost all 2D substances. Moreover, this phenomenon initiates at low temperature. At finite temperature, LTEC is negative due to out-of-plane vibrations as reported in theoretical calculation [25] of self-consistent phonons.

We have previously studied the phonon dispersion and various thermal properties like specific heat, entropy and free energy of pure graphene and how they varied with different B and N doping concentrations for graphene [26] where we also reported some estimates of LTEC for pure graphene. Since these estimates gave us negative values of thermal expansion, we now recalculate the estimates by carefully understanding the origin of negative expansion.

Therefore apart from study of thermal expansion, the primary goal of this paper has been to emphasize on the reasons for negative thermal expansion by focusing on contribution by individual phonon branches. Due to the 2-D structure of graphene with 2 atoms per unit cell, the 4 external phonon branches are usually present with 2 acoustic and 2 optic ones. The external branches arise due to intermolecular vibrations representing lattice modes. We calculate their contribution to thermal expansion separately. The other two branches in the transverse direction identified as ZA (transverse acoustic) and ZO (transverse optic) originate due to quasi-2D behavior and are very special modes of vibration for 2-D structures. Their restoring forces are derived from in plane modes and these behave very differently. Realizing that the Grüneisen parameter for ZA mode is largely negative, we have focused on its contribution separately from other branches and found that it is the most important branch which accounts for NTE in graphene. Although a detailed study for pure graphene looking into the negative thermal expansion has been made [17] but we have extended the LTEC study of graphene to doped graphene also by varying the boron and nitrogen dopant concentrations leading to h-BN sheet. Therefore despite the fact that quite a lot has been studied recently outlining the significance of ZA mode on thermal properties of graphene, this study supplements the recent study on the subject, giving detailed results of temperature dependence on thermal expansion and comparing most of the available results. Since pure graphene is not a suitable material for device applications, a study of doped graphene based on B or N will be of great practical use.

The present paper is structured as follows. After introduction in section I where graphene, the overview of its various electronic and thermodynamic properties and h-BN and its various properties are presented, a brief summary of theory and computational details are presented in Sec. II. Section III gives results and discussion which includes lattice thermo-dynamical properties like thermal expansion, surface bulk modulus and mode Grüneisen parameters as acquired from the variation of free energy with volume for pure graphene, B and N doped graphene and h-BN sheet. Sec. IV contains summary and conclusions.

## 2. THEORY AND COMPUTATIONAL DETAILS

### 2.1) Theory

To calculate the frequencies of phonon, the VASP (Vienna ab-initio Simulation Package) [27-28] code which is a density functional theory (DFT) based software was used in combination with phonopy [29] software. The VASP software is based on ab-initio density functional theory. The position coordinates of the atoms in unit cell and lattice parameters are given as input to the software with necessary conditions set by input parameters for cell relaxation. After relaxing the cell for minimum energy configuration, the lattice parameter of the structure is calculated. The structure is given small perturbations through displacements of atoms and VASP is again run under density perturbation theory for calculation of the dynamical matrix. The phonopy code uses the dynamical matrix obtained in the VASP to calculate the phonons and related thermodynamic properties. Under quasi harmonic approximation, thermal expansion gives rise to implicit shift in phonon frequencies and can be procured from dependence of the phonon frequencies on volumes, termed as Grüneisen parameter that can be defined [30,31] as

$$\gamma_{\mathbf{q}j} = -\frac{\partial \ln \omega}{\partial \ln V} = -\frac{V_0}{\omega_{\mathbf{q}j}} \frac{\partial \omega_{\mathbf{q}j}}{\partial V} \qquad (1)$$

Where $V_0$ is equilibrium volume, $\mathbf{q}$ is the phonon wave vector, j stands for different phonon branches and $\omega_{\mathbf{q}j}$'s the harmonic phonon frequencies at particular values of $\mathbf{q}$ and j.

The mode Grüneisen parameters are the key ingredient in thermal expansion mechanisms. These parameters are generally positive but for some modes and for specific q values, they are found to be negative in a narrow temperature range. They are highly negative for lowest acoustic mode in case of graphene which gives negative values for thermal expansion coefficient at low temperatures. This may be due to that fact that only lower acoustic modes can be excited at low temperatures.

In absence of external pressure, the equilibrium structure of a crystal structure can be obtained at temperature T by free energy minimization w.r.t. its degrees of freedom.

The expression for quasi-harmonic free energy is written as [30]

$$F_{qh} = \phi(V) + k_B T \sum_{\mathbf{q},j} \ln\left[2\sinh\left(\frac{\hbar\omega_{\mathbf{q}j}}{2k_B T}\right)\right] \tag{2}$$

Where $\phi(V)$ is the static lattice contribution which can be written [30] as

$$\phi(V) = \phi(V_0) + \frac{1}{2}\varepsilon^2 V_0^2 \left(\frac{\partial^2 \phi}{\partial V^2}\right) = \phi(V_0) + \frac{1}{2}\varepsilon^2 V_0 B \tag{3}$$

Where the bulk modulus B is

$$B = V_0 \left(\frac{\partial^2 \phi}{\partial V^2}\right) \tag{4}$$

In case of 2D structure, we define surface bulk modulus as

$$B_S = S\left(\frac{\partial^2 \phi}{\partial S^2}\right) = LB \tag{5}$$

Where S is the surface area and L is an arbitrary length which needs to be introduced in the otherwise 3-D standard software.

Using condition for minimum energy, expression for thermal expansion coefficient β [derived in 30] is

$$\beta = \frac{k}{V_0 B} \sum_{\mathbf{q}j} \gamma_{\mathbf{q}j} \left(\frac{\hbar\omega_{\mathbf{q}j}}{k_B T}\right)^2 \frac{e^{\hbar\omega_{\mathbf{q}j}/kT}}{(e^{\hbar\omega_{\mathbf{q}j}/kT} - 1)^2} \tag{6}$$

LTEC ($\alpha$) (for 2D materials) is defined as

$$\alpha(T) = \frac{1}{a(T)}\frac{da(T)}{d(T)} = \frac{1}{2V}\frac{\partial V}{\partial T} = \frac{1}{2}\beta \tag{7}$$

As is evident from equation (6), the thermal expansion involves summation over all branches and the mode dependent Grüneisen parameters. It has been found that mode dependent Grüneisen parameter is very sensitive to the choice of branch. Therefore we introduce $\beta_j$ for each branch to calculate branch dependent thermal expansion such that

$$\beta = \sum_j \beta_j \qquad (8)$$

**2.2) Computational parameters**

In the calculations, we have made use of the Perdew-Burke-Ernzerhof (PBE) [32] functional which is an exchange correlational functional of generalized gradient approximation (GGA). The cut off energy for plane waves was taken as 750eV. A 4x4 supercell has been employed which makes a sheet of 32 atoms to simulate the sheet of graphene. Two sheets of graphene are maintained at a separation larger than 10Å in the transverse direction in order to avoid interlayer interactions. For sampling of Brillouin zone, the Monkhorst-pack scheme is employed. A Γ centered k mesh of size 7 x 7 x 1 is used for relaxing the structure. We have made use of tetrahedron method for treating the partial occupancies with Blöchl corrections [33]. The geometry of coordinates of the system was optimized til the Hellmann-Feynman forces were as low as 0.005eV/Å. The phonon frequencies and surface bulk modulus are used in calculation of LTEC using equations (6) and (7).

## 3. RESULTS AND DISCUSSION

### 3.1) Pure Graphene

The study of pure graphene under quasiharmonic approximation gives an approximate value of Grüneisen parameters and bulk modulus which contribute in calculation of thermal expansion coefficient. We have earlier obtained the phonon dispersion for pure and B and N doped graphene and compared the dispersion curves with various theoretical and experimental results [26].

#### 3.1.1) Energy-volume Curve

Under quasi harmonic approximation, we study the free energy of the structure at different unit cell volumes and found the E-V curve which is shown in Figure 1. It is to be noted that the volume here is representative of unit cell area. This first derivative of the energy-volume curve gives the minimum unit cell volume of 63.25 Å$^3$. The lattice constant corresponding to unit cell of this volume of graphene sheet is 2.47 Å. The minimum free energy is -18.47 eV. This value of lattice constant for graphene is very close to 2.46 Å as reported in literature [19].

Thus bulk modulus using equation (4) is obtained from second derivative of the fitted curve to the calculated E-V data as shown in Figure 1. Hence we obtained surface bulk modulus as defined by us (equation 5) for pure graphene to be equal to 90.24 N/m. The bulk modulus of graphene has earlier been studied by many authors [34-37]. Reich et al. [34] have reported a bulk modulus value of 700GPa and Milowaska et. al. [37] has reported a value of 528 GPa for pure graphene. It may be noted that for calculation of volume of the unit cell they take thickness of the cell around 3.4 Å. The 2D bulk modulus as defined by Kalosakas et. al. [35] has been reported to be 200N/m for graphene. We have found a bulk modulus value of 262 GPa which is in qualitative agreement when compared to previous studies based on similar cell volume calculations.

### 3.1.2) Mode dependent Grüneisen parameter

The Grüneisen parameter plays an important role in finding the dependence of various thermodynamic properties of materials on temperature ranging from phonon frequency shift to thermal expansion of a material. Generally these parameters are positive, as phonon frequencies diminishes when a solid expands, although some Grüneisen parameters are reported to be negative for low frequency acoustic modes and usually combine with positive values from different modes in the end diluting the effectiveness of positive Grüneisen parameters. Interestingly, thermal expansion can not only be diluted by these but also turn positive expansion to a negative expansion. To analyze thermal contraction, we have shown in Figure 2, mode dependent Grüneisen parameters of graphene. These are obtained from phonon frequencies by interpolation with a polynomial and calculated for the ground state configuration. Interestingly some bands in graphene (lowest transverse acoustic mode) show large negative Grüneisen parameters, as compared with other branches. Thus at very low temperatures when optical modes are still not excited whose Grüneisen are positive, the contribution from the negative Grüneisen parameters will highly contribute to negative thermal expansion. The Grüneisen parameters corresponding to the lowest transversal acoustic ZA modes are negative. These negative Grüneisen parameters will correspond to negative thermal expansion in graphene.

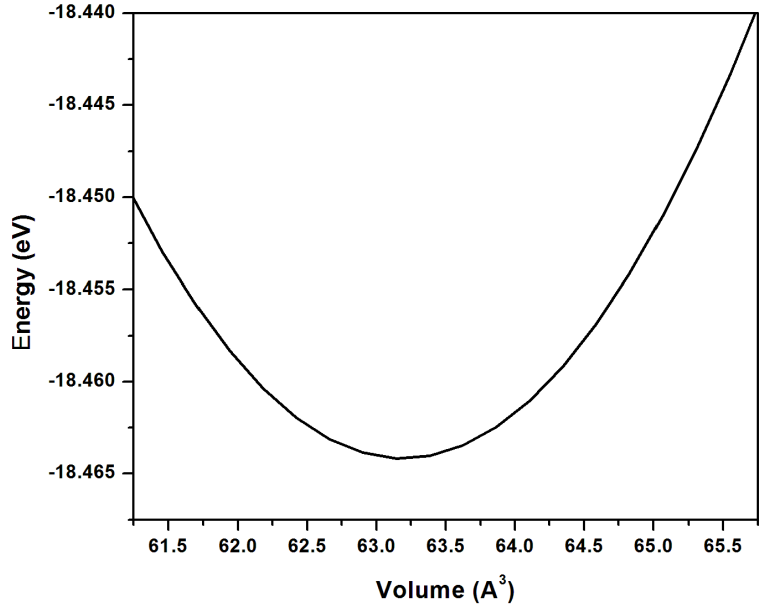

Figure 1 Energy-Volume Curve for pure graphene where volume is area multiplied by a constant height in z-direction

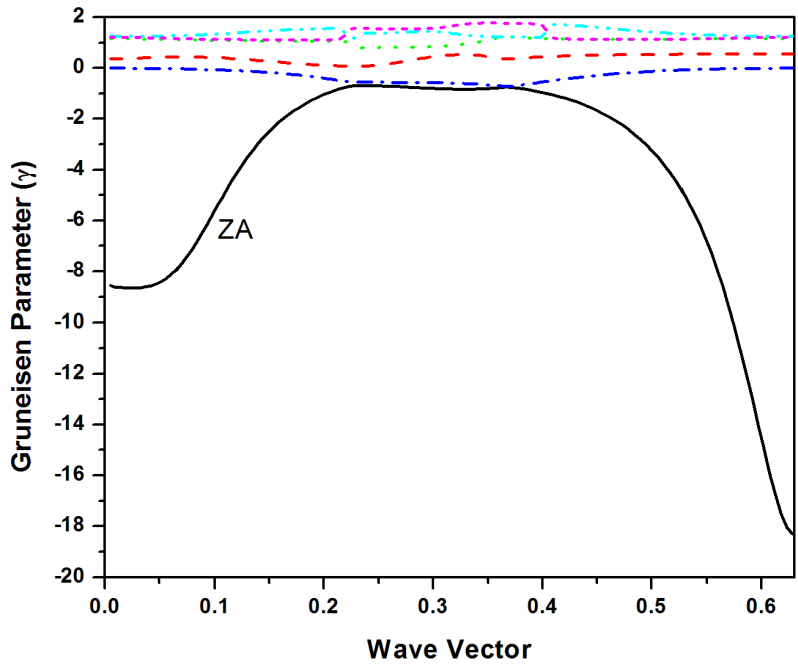

Figure 2 Grüneisen parameters for various phonon branches of graphene

### 3.1.3) Thermal Expansion Coefficient

An important feature of thermal properties of graphene is their extremely high thermal conductivity, which makes it a suitable candidate for use as a heat control device in high speed integrated circuits. Exact knowledge of the temperature dependence of linear thermal expansion coefficient (LTEC) is also important for such applications. Many authors have contributed to the calculation of TEC using various theoretical models [10-17] and experimental techniques [18-20]. Bao et al. [18] have measured experimentally the coefficient of thermal expansion in a very narrow range of temperature (300K– 400 K) by examining the variation in the sagging of a suspended graphene piece and found it to be negative only up to ~350 K with a RT value of -7 X 10-6 K-1 . Mounet et al. [10] estimated the LTEC of graphene using ab initio DFT calculation and estimated that the LTEC of graphene is negative in the whole temperature range under study (0-2500K). Michel et al. [17] have studied thermal expansion using anharmonic force constants and estimated the RT value of LTEC to be -1.6X10-7K-1 which is very much off the experimental values [18-20].

As observed in figure 3a, there is a large variation between theoretical and experimental data. Even the experimental data shows variation. In the low temperature regime, the size dependence might have a role to play [17] -smaller size tends to reduce the large negative value but it has not been mentioned by experimentalists. Our results are in overall agreement with those of Mounet et al.[10] and show a better agreement with experiments above room temperature.

To get complete and accurate understanding of the behavior of LTEC near room temperature, which is important in designing graphene-based heat devices, we studied the branch dependent LTEC in detail.

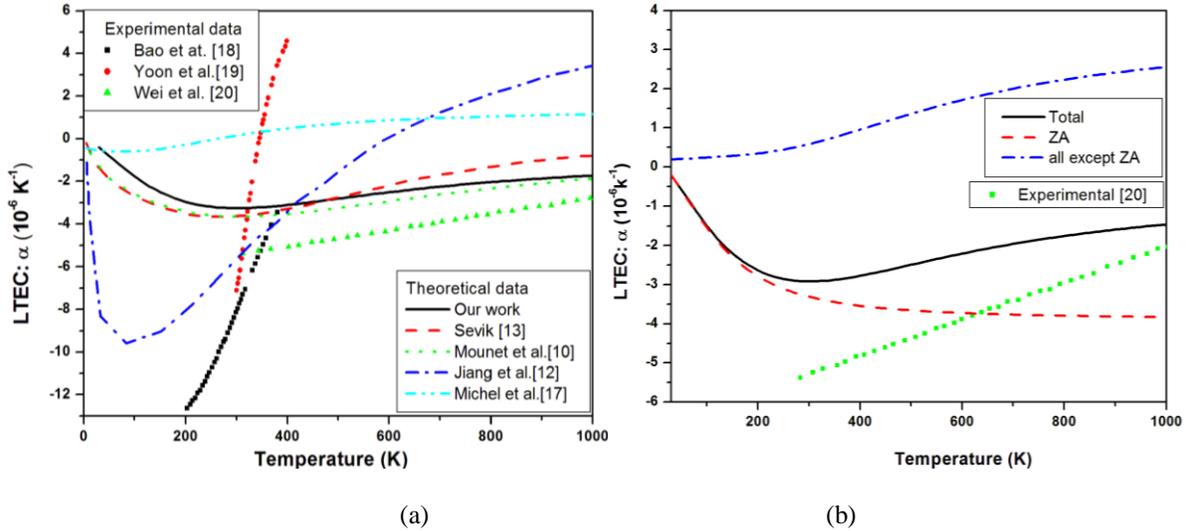

(a)          (b)

Figure 3(a) Theoretical linear thermal expansion coefficient α of single layer graphene plotted as a function of temperature and compared with experimental data [18] (black squares), [19] (red circles) and [20] (green triangles) and theoretical measurements [10] (green dots), [13] (red dash line), [12] (blue dash dot line) and [17] (cyan dash dot dot line). (b) Branch dependent thermal expansion coefficient (Total-black solid line, ZA mode contribution – red dash line and all except ZA contribution- blue dash dot line) compared with experimental data [20] shown by green squares in the curve.

DFPT calculations of pure graphene for LTEC calculations show a negative behavior in whole range of temperature under study with a RT value of (-3.26X10$^{-6}$). In this calculation we have made use of surface bulk modulus calculated earlier using E-V curve. Our results matches well with the theoretical studies [10, 13] as shown in figure 3(a) but there is still a discrepancy as compared to experimental results. Theoretical measurements by [12] are very close and [17] are very much off to experimental value of LTEC. The reasons for this disagreement have been attributed to finite size effect incorporated by [17]. To analyze it further, we present partial contributions to LTEC arising from each branch separately as the phonon branch ZA in particular has large negative Grüneisen parameters and must be visualized separately. Figure 3(b) shows the branch dependent LTEC which clearly chows that thermal expansion of graphene is negative only due to presence of large negative Grüneisen parameters corresponding to ZA branch. Here we have made use of equation (8) to calculate branch dependent thermal expansion. The dominant contribution to thermal expansion in graphene is from ZA branch which is negative. The overall thermal expansion is negative due to dominant effect of ZA branch compared to other branches. The experimental data reported here is from three sources [18-20] –all obtaining it from the shift in 2D peak and exploiting thermal expansion as responsible for the shift. However in the small domain of overlap of three measurements i.e. around 300- 400K, there is no agreement between the three. In the high temperature limit above RT our results show a reasonably good agreement with the experimental data. It may be possible that grüneisen parameter depends on temperature but there is no theoretical or experimental estimate of dependence of grüneisen parameters on temperature. This assumption introduces uncertainty in estimation of thermal expansion.

### 3.2) Doped graphene

Since the B (N) doped structures are stable upto 25% doping [26], we have done the quasiharmonic study for two cases only i.e. 12.5% and 25% The atomic configurations corresponding to doped sheet is shown in figure 4.

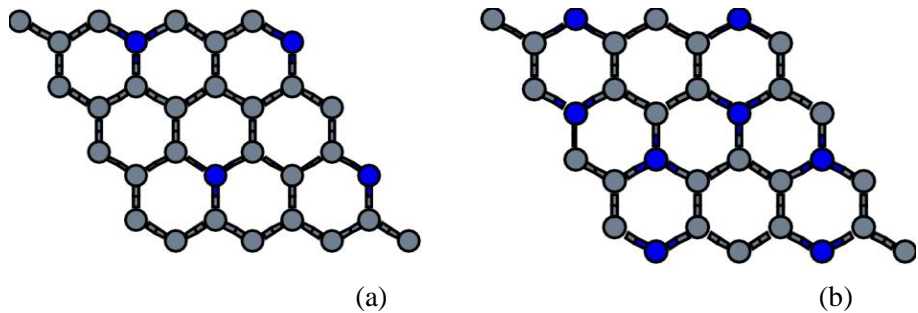

(a)  (b)

Figure 4 Schematic diagrams of (a) 12.5% and (b) 25% doping concentrations

We have also considered the minimum doping configuration. The minimum doping can be one atom in a sheet of 32 atoms i.e. 3.125% in this case. But the configuration with low doping i.e. 1/2/3 atoms in a sheet is also thermodynamically unstable where almost all of the ZA mode frequencies are imaginary for boron doping and half of ZA frequencies are imaginary for nitrogen doping, as doping of B atoms introduces more lattice mismatch compared to N atoms doping which makes the structure thermodynamically more unstable. Thus considering one, two or three atoms doping in a hexagonal ring gives the thermodynamically unstable structures. We finally studied thermal expansion in only above two thermodynamically stable structures as already reported in our previous studies [26].

### 3.2.1) N-Doping

Thermal expansion coefficient has been calculated for the doped structure for different configurations and compared with that of pure graphene as shown in figure 5. There is lowering of LTEC value with increasing doping concentration with a RT value of $-8.19 \times 10^{-6}$ for 12.5% doping and $-9.66 \times 10^{-6}$ for 25% doping. The large increase in negative value of thermal expansion with doping is primarily concerned with the ZA branch (similar to that shown earlier for pure graphene in figure 3(b)) which has more negative Grüneisen parameters near Γ point as low as -150 and -200 for 12.5% and 25% N doped sheets respectively. The surface bulk modulus values of doped sheets are also depicted using E-V curve to be 106.32N/m and 97.2 N/m for 12.5% and 25% N doped sheets respectively. It has been shown in an earlier research paper that bulk modulus value of N doped graphene increases initially with doping and then tend to decrease as doping is increased further [37]. This is apparently due to decreased cell volume on doping. At higher doping concentrations the modified interactions also play a role.

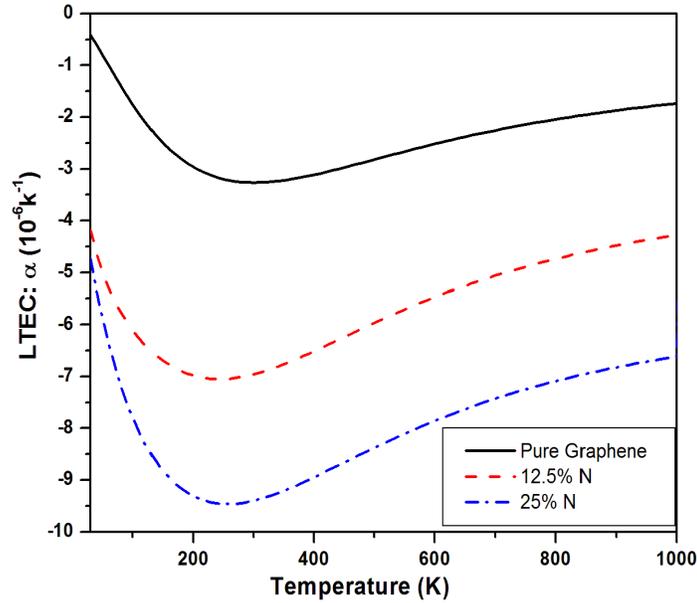

Figure 5 LTEC of 12.5% and 25% N-doped sheet compared with pure graphene case

### 3.2.2) B-Doping

Thermal expansion coefficient for the B doped structure for different doping configurations is shown in figure 6. There is lowering of thermal expansion value with doping with a RT value of $-9.17 \times 10^{-6}$ for 12.5% doping and $-11.16 \times 10^{-6}$ for 25% doping. The increase in negative value of thermal expansion with B doping is comparatively larger than N doping although the trend is similar in both cases. The contribution of ZA branch in thermal expansion is to make it negative as the Grüneisen parameters for ZA branch for the B doped structures are as low as -105 and -150 for 12.5% and 25% respectively. However the increase in negative value of LTEC for the case of B doped sheet as compared to N doped sheet is much larger due to decrease in its surface bulk modulus value. The surface bulk modulus values of B doped sheets are depicted to be 92.5 N/m and 73.5 N/m for 12.5% and 25% doping configurations respectively.

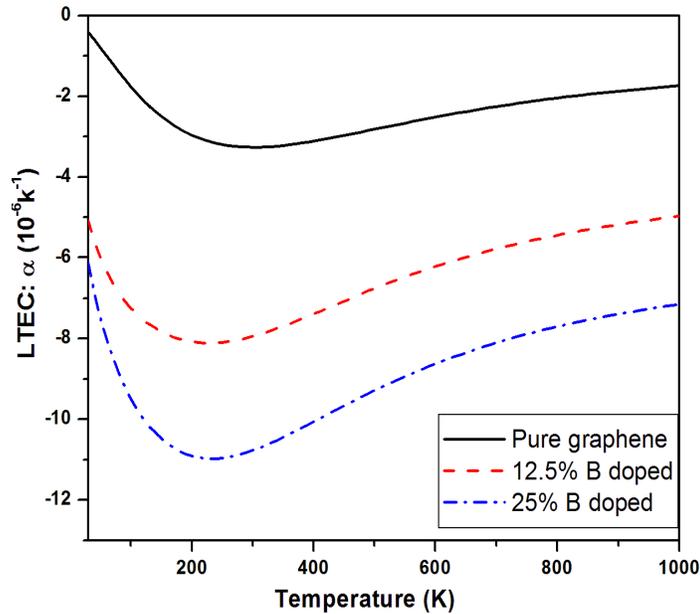

Figure 6  LTEC of 12.5% and 25% B-doped sheet compared with pure graphene case

### 3.3) h-BN sheet

To understand thermodynamic properties of h-BN sheet, the first step is to inspect lattice vibrational modes (phonons).

### 3.3.1) Phonon Dispersion

Similar to phonon dispersion of graphene calculated earlier [26], h-BN sheet also have 6 phonon branches (three acoustic (A) and 3N-3 optical (O) phonon modes).

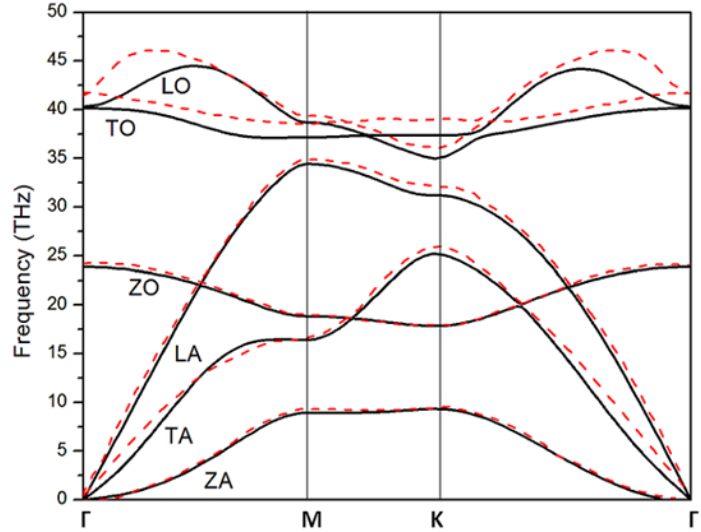

Figure 7 Phonon Dispersion for h-BN (solid lines) compared to that in Ref [38] (red dashed lines). The longitudinal, transversal and out-of-plane acoustical/optical modes are represented by LA /LO, TA /TO and ZA /ZO respectively**.**

The phonon frequencies are obtained under harmonic approximation which gives phonon dispersion curve of h-BN sheet as shown in figure 7. The curves have been obtained by simulating the h-BN sheet over a set of points on the line for different directions between Γ points. We have made use of density functional perturbation theory with displacement along all the three directions which give rise to 6 phonon branches in dispersion curve. The phonon dispersion curve is in good agreement with reference [38] where Wirtz *et al.* have studied the phonon dispersion curve using molecular dynamics. Our dispersion curve also matches with [39] where vibrational properties of h-BN are studied. They have shown that in a strictly 2D material, the electrostatic interactions do not produce any macroscopic term at Γ point and hence no splitting is observed between LO/TO modes in an h-BN sheet. Our phonon dispersion curve is also fairly in agreement with experimental studies [40, 41] done using inelastic x-ray scattering and Raman analysis methods.

### 3.3.2) Energy volume curve

Similar to the pure graphene, the E-V curve obtained for h-BN sheet is shown in figure 8. The volume (area of sheet multiplied by a constant height along z-axis) which minimizes the curve is 55.26 Å$^3$.

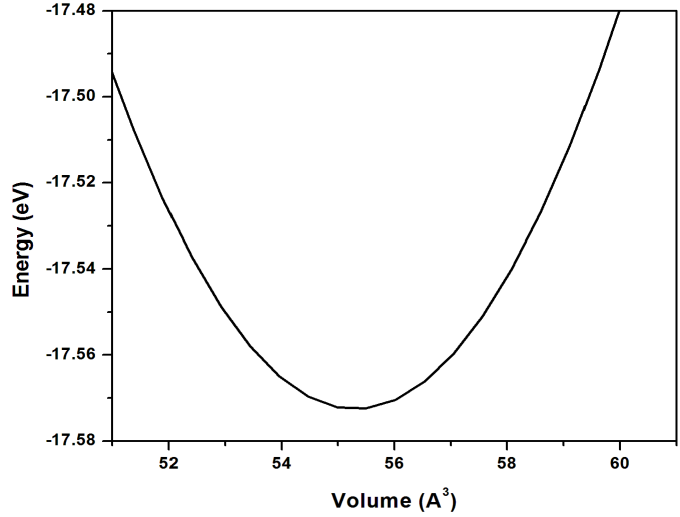

Figure 8 Energy-Volume Curve for h-BN sheet

The lattice constant corresponding to minimum unit cell volume is 2.515 Å. The surface bulk modulus found to be 74.6N/m for h-BN sheet which is much smaller than surface bulk modulus of pure graphene.

### 3.3.3) Grüneisen parameter

The mode dependent Grüneisen parameter for h-BN sheet has been obtained for various phonon branches corresponding to different q-values and shown in figure 9(a). The Grüneisen parameter corresponding to lowest transverse acoustic mode is negative in the whole region with lowest values approaching as low as – 230 larger than doped and pure graphene case as shown in figure 9(b). There is a gap between ZA mode Grüneisen parameters and other acoustic and optical mode Grüneisen parameters contrary to the case of graphene as shown in figure 2 where there is no gap. This observation is similar to the observation of difference in phonon dispersion curve of graphene and h-BN.

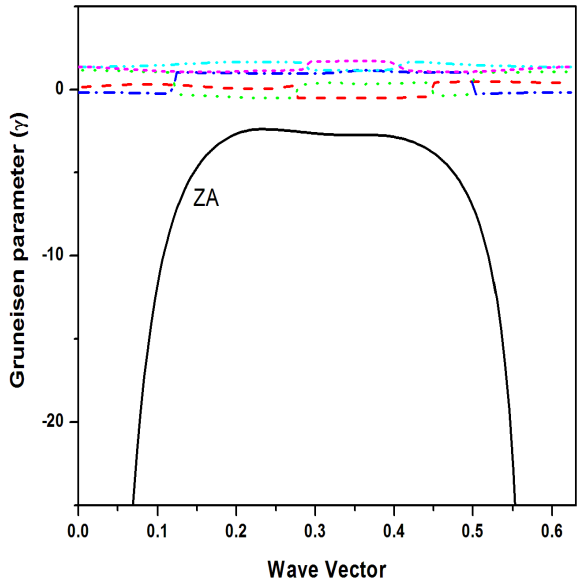

(a)

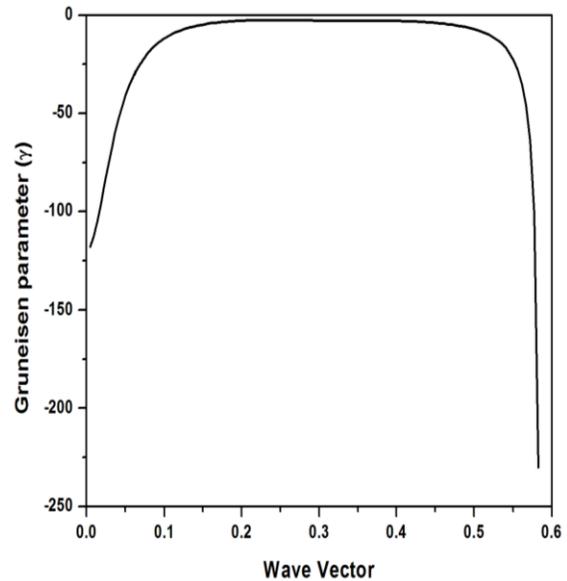

(b)

Figure 9(a) Mode dependent Grüneisen parameters, (b) Grüneisen parameter of ZA mode

### 3.3.4) Thermal Expansion Coefficient

For calculating LTEC of single layer h-BN sheet, several authors have used various methods [13, 24, 42-44] using various experimental techniques [42-44] and theoretical models [13, 24]. Sevik [13] theoretically estimated the LTEC in the temperature range of 0-1500 K by doing calculations using DFPT as implemented in VASP software. Thomas *et al.* [24] have investigated LTEC in a huge range of temperature using molecular dynamics. The results of various experimental techniques for bulk h-BN crystals [40-41] are in the temperature range 0-400K and lower than the theoretical single sheet estimates given by Sevik [13].

We obtained the behavior of LTEC for h-BN sheet using DFPT under QHA as shown in Figure 10(a). Although the RT value of LTEC is much larger than that determined by Sevik [13] of the order of $-12.86 \times 10^{-6}$ but the behavior is in conformity with B and N doped structures. The negative thermal expansion can be explained by The existence of more negative Grüneisen parameters corresponding to the transverse acoustic (ZA) mode and lower surface bulk modulus compared to graphene explains the occurrence of negative thermal expansion. The experimentally measured values in the temperature range 0–400 K are for bulk h-BN crystals which are lower than estimates for single h-BN sheet. The prominent difference in estimates arises due to the strong anharmonic character of h-BN structure, which is not taken into account in QHA calculations. Alternately, the LTEC values predicted by Thomas et al. [24] based on fully atomistic simulations gives a lower estimate compared with Sevik [13]. Figure 10(b) shows branch dependent LTEC which depicts the contribution of ZA branch is of utmost importance to total thermal expansion as compared to other branches.

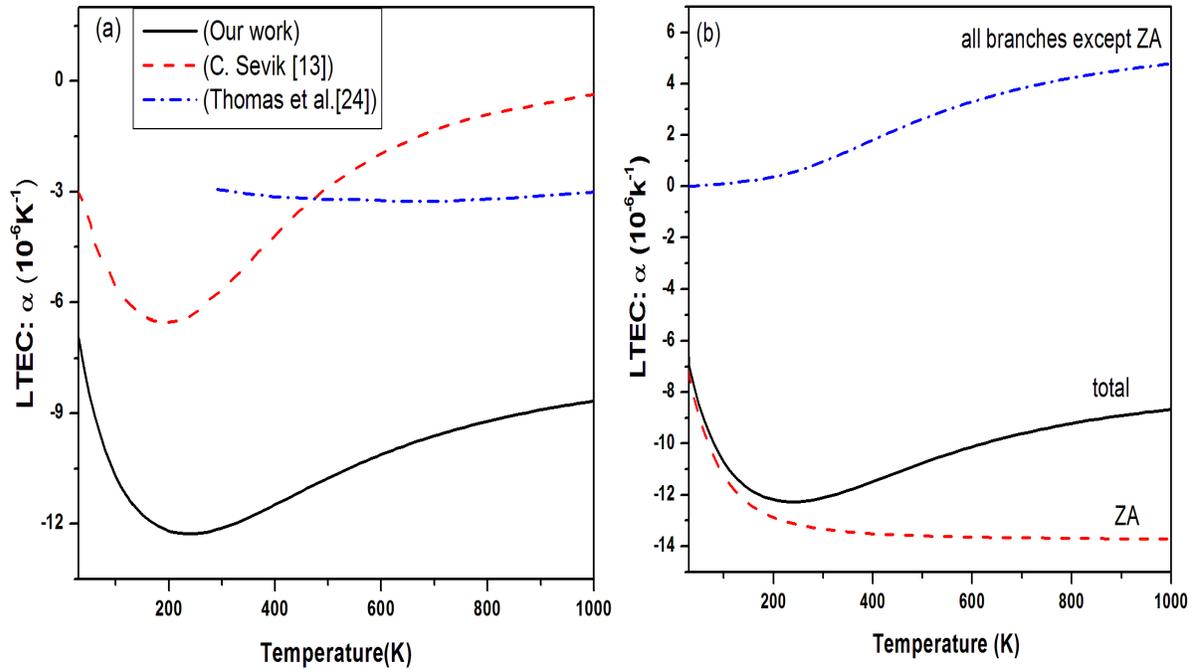

Figure 10(a) Temperature dependence of LTEC for single layer *h*-BN black solid line. Our results are compared with various theoretical measurement (Ref [13] red dashed line and Ref [24] blue dash dot line). (b) Branch dependent thermal expansion

Figure 11 shows a comparison of LTEC for pure and doped graphene compared with h-BN sheet. The graph trend shows an increase in negative thermal expansion with increase in N and B doping and further increases in h-BN sheet. We also observe that B doping makes the LTEC value more negative as compared to same doping percentage of N atom. Further we notice that our estimates are overestimating the LTEC values for h-BN sheet when compared to [13] by a factor close to 2. However, our study of systematic doping by increasing B and N concentrations follows the pattern of LTEC as shown in figure 11 quite well where we present results for doping of graphene with individual B and N atoms going from 12.5% to 25% and finally 50% B and N doping together creating h-BN sheet.

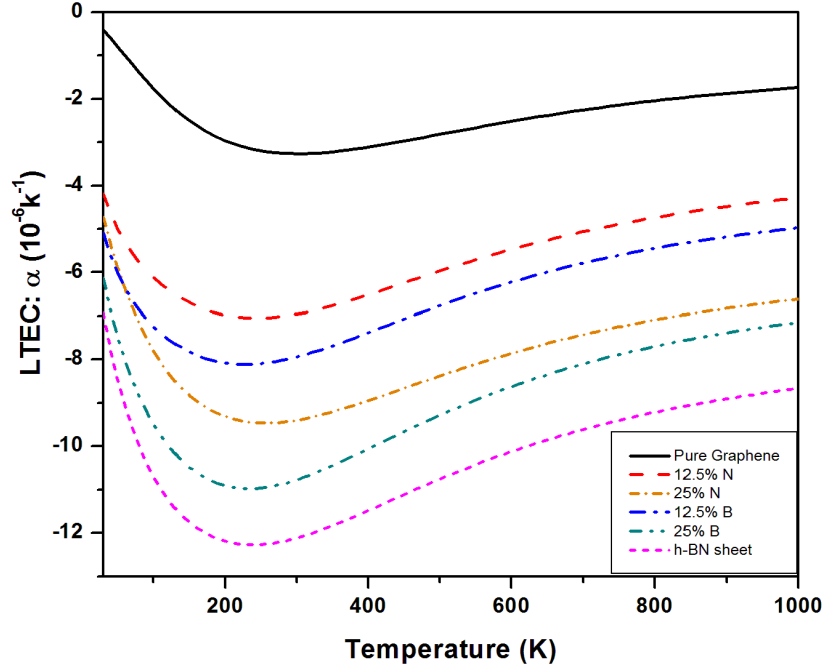

Figure 11 Temperature dependence of LTEC for pure graphene compared with B and N doped graphene and single layer *h*-BN sheet

## 4) SUMMARY AND CONCLUSIONS

In this paper, we have made a detailed first-principles study using the quasi-harmonic approximation at the GGA-PBE level to obtain the finite temperature dependence of LTEC in pure graphene. The thermal expansion coefficient is negative in the whole temperature region under study and is in qualitatively good agreement with available theoretical data although it deviates from experimental studies especially in the low temperature region. In the high temperature region our results are in better agreement with experiments as compared to some other calculations.

Thereafter we have extended the calculation to doped graphene. We have already shown that Boron and nitrogen doped graphene is an interesting material whose band gap and other electronic properties can be designed by varying the concentration and type of the dopant [6, 8]. The calculation of thermal properties is thus of great importance for using these doped materials in practical applications.

We notice that there is consistent trend of increase in negative Grüneisen parameters with B and N doping which continues even for h-BN sheet. The values of grüneisen parameters tend to vary by a factor of 5 or more for B or N doping upto 25%. The same concentration of B atom doping brings about larger change in the negative LTEC value as compared to N doping. The results for thermal expansion presented here for doped graphene sheet are new.

We extend this approach further by substituting all the carbon atoms by B and N atoms to obtain h-BN sheets. We have presented a study of *h*-BN phonon dispersion and estimated its linear thermal expansion coefficient similarly. The phonon frequencies and phonon dispersions are in good agreement with various theoretical and experimental results. Some calculations report results for thermal expansion of h-BN sheets. Although our results follow similar qualitative temperature dependence, the results differ by about a factor of two. Our results (figure 11) follow a pattern showing a systematic increase in thermal expansion values as the concentration of dopants increases.

On the basis of the results acquired here for various pure and different doping concentrations of 2D graphene it emerges that the negative thermal expansion originates from ZA modes. These modes result in highly negative Grüneisen parameters and their contribution dominates from the other planar modes. The planar modes (LA, LO, TA, TO) and the remaining transverse modes (ZO) contribute positively but their combined effect remains systematically smaller than the negative contribution from ZA mode. It seems that this feature will always be a unique feature of all 2D materials. Exceptions could be those 2D materials like $MoS_2$ and $MoSe_2$ having larger number of atoms in unit cell and many more branches in planner directions contributing positively to thermal expansion which may end up in combining to nullify most of the negative contribution from ZA mode as measured by Sevik [13].

Experimental measurements for thermal expansion for graphene lack consistency. There is significant variation and unusual temperature dependence in low temperature region. More experimental work for pure graphene as well as for doped graphene for thermal expansion would be useful to correlate theoretical results.

## 5) ACKNOWLEDGEMENTS

We are grateful to VASP and phonopy team for providing the code, the departmental computing facilities at Department of Physics, Panjab University, Chandigarh and the HPC facilities at IUAC (New Delhi). VKJ acknowledges Alexander von Humboldt Foundation (Germany) for support as a Guest Professor at University of Würzburg.

## REFERENCES

1. Novoselov KS, Geim AK, Morozov SV and Jiang D, et al, Electric field in atomically thin carbon films, Science 2004 Oct 22;306(5696):666-9.
2. Geim AK and Novoselov KS, The rise of graphene, Nature Materials 2007 Mar: 6(3):183–91.
3. Geim AK, Graphene: status and prospects, Science 2009 Jun 19; 324(5934):1530-4.

4. Castro Neto AH, Guinea F, Peres NMR, Novoselov KS and Geim AK, The electronic properties of graphene, Rev. Mod. Phys. 2009 Jan 14; 81:109-62.
5. Avouris P, Chen Z and Perebeinos V, Carbon based electronics, Nat. Nanotechnol. 2007 Oct;2(10):605-15.
6. Rani P and Jindal VK, Designing band gap of graphene by B and N dopants, RSC Adv.2013;3(3):802-12.
7. Rani P and Jindal VK, Stability and electronic properties of isomers of b/n co-doped graphene, Appl Nanosci. 2014 Nov;4(8):989–96.
8. Rani P, Dubey GS and Jindal VK, Physica, DFT study of optical properties of pure and doped graphene, Physica E 2014 aug 31;62;28–35.
9. Nika DL, Pokatilov EP, Askerov AS and Balandin AA, Phonon thermal conduction in graphene: Role of Umklapp and edge roughness scattering, 2009 April; 79(15):155413.
10. Mounet N and Marzari N, First-principles determination of the structural, vibrational and thermodynamic properties of diamond, graphite, and derivatives, Phys. Rev.2005 May 31;71(20); 205214.
11. Zakharchenko KV, Katsnelson MI and Fasolino A, Finite temperature lattice properties of graphene beyond the quasiharmonic approximation, Phys. Rev. Lett. 2009 Jan 29; 102: 046808.
12. Jiang JW, Wang JS and Li B, Young's modulus of graphene: A molecular dynamics study, Phys. Rev. B 2009 Sep 23; 80: 205429.
13. Sevik C, Assessment on lattice thermal properties of two-dimensional honeycomb structures: Graphene, h-BN, h-MoS2, and h-MoSe2, PRB 2014, Jan 21; 89:035422.
14. Kim CW, Kang SH and Kwon YK, Rigid unit modes in sp−sp2 hybridized carbon systems: Origin of negative thermal expansion, Phys. Rev. B 2015;92:245434.
15. Jiang JW, Wang BS, Wang JS and Park HS, A review on flexural mode of graphene: lattice dynamics, thermal conduction, thermal expansion, elasticity, and nanomechanical resonance, J. of Phys. Cond. Matter 2015; 27(8):083001.
16. Michel KH, Costamagna S and Peeters FM, Theory of thermal expansion in 2D crystals, Physica status solidi (b) 2015 Aug 6; 252 (11):2433-37.
17. Michel KH, Costamagna S and Peeters FM, Theory of anharmonic phonons in two-dimensional crystals, Phys. Rev. B 2015; 91(13):134302.
18. Bao W, Miao F, Chen Z et. al., Controlled ripple texturing of suspended graphene and ultrathin graphite membranes Nat. Nanotechnol. 2009; 4:562.
19. Yoon D, Son YW and Cheong H, Negative Thermal Expansion Coefficient of Graphene Measured by Raman Spectroscopy, Nano Lett.2011;11(8):3227-31.

20. Pan W, Xiao J, Zhu J, Yu C et al., Biaxial Compressive Strain Engineering in Graphene/Boron Nitride Heterostructures, Sci. Rep. 2012;2:893.

21. Blase X, Rubio A, Louie SG, and Cohen ML, Quasiparticle band structure of bulk hexagonal boron nitride and related systems, Phys. Rev. B 1995; 51:6868.

22. Watanabe K, Taniguchi T, and Kanda H, Direct-bandgap properties and evidence for ultraviolet lasing of hexagonal boron nitride single crystal, Nat. Mater. 2004 May;3:404-9.

23. Golberg D, Bando Y, Huang Y, Terao T, Mitome M, Tang C and Zhi C, Boron nitride nanotubes and nanosheets, ACS Nano 2010 Jun 22;4(6):2979-93.

24. Thomas S, Ajith KM, Chandra S and Valsakumar MC, Temperature dependent structural properties and bending rigidity of pristine and defective hexagonal boron nitride, J Phys Condens Matter. 2015 Aug 12;27(31):315302.

25. hu Y, Chen J and Wang B, On the intrinsic ripples and negative thermal expansion of graphene, Carbon 2015 Dec; 95:239-49.

26. Mann S, Rani P, Kumar R, Dubey GS and Jindal VK, Thermodynamic properties of pure and doped (B, N) graphene, RSC Adv. 2016;6(15):12158-68.

27. Kresse G and Furthmüller J, Efficient iterative schemes for ab initio total-energy calculations using a plane-wave basis set, Phys. Rev. B 1996 Oct 15; 54(16):11169.

28. Kresse G and Joubert D, From ultrasoft pseudopotentials to the projector augmented-wave method, Phys. Rev. B 1999 Jan 15; 59(3):1758.

29. Togo A, Oba F and Tanaka I, First-principles calculations of the ferroelastic transition between rutile-type and CaCl2-type SiO2 at high pressures, Phys. Rev. B 2008 Oct;78(13):134106.

30. Jindal VK and Kalus J, Calculation of thermal expansion and phonon frequency shift in deuterated naphthalene. Phys. Stat. Sol. (b) 1986; 133:189.

31. Bhandari R and Jindal VK, Calculation of thermal expansion and implicit phonon frequency shift in deuterated anthracene, J. Phys.: Condens. Matter 1991; 3:899-907.

32. Perdew JP, Burke K and Ernzerhof M, Generalized Gradient Approximation Made Simple, Phys. Rev. Lett.1996;77(18):3865-68.

33. Blöchl PE, Projector augmented-wave method, Phys. Rev. B 1994 Dec 15; 50(24):17953.

34. Reich S and Thomsen C, Elastic properties of carbon nanotubes under hydrostatic pressure, Phys. Rev. B 2002; 65:153407.

35. Kalosakas G, Lathiotakis NN, Galiotis C and Papagelis K, In-plane force fields and elastic properties of graphene, J. Appl. Phys.2013;113:134307.

36. Zakharchenko KV, Katsnelson MI, Fasolino A, Finite temperature lattice properties of graphene beyond the quasiharmonic approximation, Phys. Rev. Lett. 2009; 102(4):046808.

37. Milowska KZ, Woińska M and Wierzbowska M, Contrasting Elastic Properties of Heavily B-and N-doped Graphene with Random Impurity Distributions Including Aggregates, Journal of Physical Chemistry C 2013;117 (39):20229-20235.
38. Wirtz L and Rubio A, Ab initio calculations of the lattice dynamics of boron nitride nanotubes, Phy. Rev. B 2003 July 30; 68(4):045425.
39. Portal DS and Hernandez E, Vibrational properties of single-wall nanotubes and monolayers of hexagonal BN, Phy. Rev. B 2002 Dec 20; 66(23):235415.
40. Serrano J, Bosak A, Arenal R, Krisch M, Watanabe K, Taniguchi T et. al., Vibrational Properties of Hexagonal Boron Nitride: Inelastic X-Ray Scattering and Ab Initio Calculations, Phys. Rev. Lett. 2007 March 1; 98:095503.
41. Reich S, Ferrari AC, Arenal R, Loiseau A, Bello I and Robertson J, Resonant Raman scattering in cubic and hexagonal boron nitride, Phys. Rev. B 2005;71:205201.
42. Belenkii GL, Salaev EY, Suleimanov RA, Abdullaev NA and Shteinshraiber V, The nature of negative linear expansion in layer crystals C, Bn, GaS, GaSe and InSe, Solid State Commun. 1985; 53(11):967-71.
43. Paszkowicz W, Pelka J, Knapp M, Szyszko T and Podsiadlo S, Lattice parameters and anisotropic thermal expansion of hexagonal boron nitride in the 10–297.5 K temperature range, Appl. Phys. A 2002 Sep; 75(3):431-35.
44. Yates B, Overy M and Pirgon O, The anisotropic thermal expansion of boron nitride Philos. Mag.1975; 32(4):847-57.